\documentclass[preprint,12pt]{elsarticle}
\usepackage{natbib}
\usepackage{graphicx}
\usepackage{mathtools}
\usepackage{amsthm}
\usepackage{amsmath}
\usepackage{amssymb}
\usepackage{bm}
\usepackage{bbm}
\usepackage{upgreek}
\usepackage[colorlinks=true,linkcolor=blue,citecolor=blue]{hyperref}
\usepackage{enumitem}
\usepackage{multirow}
\usepackage{mathrsfs}
\usepackage{adjustbox}

\DeclareMathOperator*{\argmin}{arg\,min}
\usepackage{comment}
\usepackage{color}
\usepackage{graphicx}

\usepackage{float}
\restylefloat{table}
\usepackage{caption}

\newtheorem{assumption}{Assumption}

\newtheorem{proposition}{Proposition}
\newtheorem{lemma}{Lemma}

\newcommand{\indep}{\raisebox{0.05em}{\rotatebox[origin=c]{90}{$\models$}}}

\def\eop
{\hfill$\Box$\\
}

\journal{""}
\begin{document}
\begin{frontmatter}
\title{On expectile-assisted inverse regression estimation for sufficient dimension reduction}
\author{Abdul-Nasah Soale}
\author{Yuexiao Dong\corref{cor1}}
\address{Department of Statistical Science, Temple University,
Philadelphia, PA, US, 19122}
 \ead{ydong@temple.edu}
 \cortext[cor1]{Corresponding author.}

\begin{abstract}
Moment-based sufficient dimension reduction methods such as sliced inverse regression may not work well in the presence of heteroscedasticity. 
We propose to first estimate the expectiles through kernel expectile regression, and then carry out dimension reduction based on random projections
of the regression expectiles. Several popular inverse regression methods in the literature are extended under this general framework. The proposed expectile-assisted methods outperform existing moment-based dimension reduction methods in both numerical studies and an analysis of the Big Mac data. 
\end{abstract}
\end{frontmatter}

\section{Introduction}

Since its inception about three decades ago, sufficient dimension reduction (Li, 1991; Cook, 1998a) has become a very important tool for modern multivariate analysis. For  predictor $\bm{X} \in \mathbb{R}^p$
and response $Y \in \mathbb{R}$, the goal of sufficient dimension reduction is to find  $\bm{B} \in \mathbb{R}^{p \times d}$ with $d \leq p$ such that
\begin{align}
    Y \indep \bm{X} | \bm{B}^\top \bm{X},
    \label{independence}
\end{align}
where $\indep$ means statistical independence.
The column space of $\bm{B}$ satisfying (\ref{independence}) is known as a dimension reduction space.
Under mild conditions,  Yin, Li and Cook (2008) showed that
the intersection of all dimension reduction spaces is still a dimension reduction space, and it is referred to as the central space for the regression
 $Y$ on $\bm{X}$. We denote the central space by $\mathcal{S}_{Y \mid \bm{X}}$. The dimension of the central space is known as the structural dimension. 

There are many sufficient dimension reduction methods in the literature. Moment-based estimators include sliced inverse regression (SIR) (Li, 1991),
sliced average variance estimation (SAVE) (Cook and Weisberg, 1991), principal Hessian directions (Li, 1992; Cook, 1998b), sliced average third-moment estimation (Yin and Cook, 2003), and SIR-$\alpha$ (Saracco, 2005). Semiparametric estimators include minimum average variance estimation (MAVE) (Xia et al., 2002), and semiparametric dimension reduction  (Ma and Zhu, 2012; Luo, Li and Yin, 2014). Sparse dimension reduction estimators include sparse SIR (Li, 2007; Tan, Shi and Yu, 2020), sparse MAVE (Wang and Yin, 2008), coordinate-independent sparse estimation (Chen, Zou and Cook, 2010), and sparse semiparametric estimation (Yu et al., 2013). Other sufficient dimension reduction methods include ensemble sufficient dimension reduction (Yin and Li, 2011), nonlinear sufficient dimension reduction (Li, Artemiou and Li, 2011;
Lee, Li and Chiaromonte, 2013), groupwise sufficient dimension reduction (Li, Li and Zhu, 2010; Guo et al., 2015),  post dimension reduction inference (Kim et al., 2020), and online sufficient dimension reduction (Chavent et al., 2014; Cai, Li and Zhu, 2020). For general reviews, one can refer to Cook (2007),  Ma and Zhu (2013), and Dong (2020). An excellent reference is the recent book by Li (2018).

Due to their ease of implementation, SIR and SAVE are two of the most popular sufficient dimension reduction methods.
One well-known limitation of SIR and SAVE is that
they are not very efficient in the presence of heteroscedasticity.
Quantile-based methods are proposed by Wang, Shin and Wu (2018) and  Kim, Wu and Shin (2019) to address this limitation, and their proposals work better than SIR or SAVE with heteroscedastic error. However, another well-known limitation of SIR and SAVE is that they may be sensitive to specific link functions between the response and the predictor. In particular, SIR does not work well when the link function is symmetric, and SAVE is not efficient with monotone link functions. Since the quantile-based methods are extensions of SIR and SAVE, they inherit the limitation of their moment-based counterparts and may still have uneven performances with various link functions.

We propose expectile-assisted inverse regression in this paper. Our contribution is two-fold. First, we provide a general framework to extend moment-based dimension reduction methods to their expectile-based counterparts, such as expectile-assisted SIR, expectile-assisted SAVE, and expectile-assisted directional regression. Similar to the quantile-based methods, our expectile-based proposals utilize the information across different levels of the conditional distribution of $Y$ given $\bm{X}$, and perform better than the corresponding moment-based methods in the presence of heteroscedasticity. Since directional regression (Li and Wang, 2007) is known to perform well for a wide range of link functions, the expectile-assisted directional regression enjoys the additional benefit that it is no longer sensitive to the specific forms of the unknown link functions. Furthermore, to combine the information across different quantile levels, existing
quantile-based methods such as quantile-slicing mean estimation and quantile-slicing variance estimation (Kim, Wu and Shin, 2019) rely on intricate weights, and it is not clear how the choice of different weights may affect the final estimation. We propose to combine the information across different expectile levels through random projection, which has roots in the projected resampling approach for multiple response sufficient dimension reduction (Li, Wen and Zhu, 2008). Our proposed expectile-assisted estimators outperform existing methods in both simulation studies and a real data analysis. 

The rest of the paper is organized as follows. In Sections 2 and 3, we provide the population level and the sample level development of expectile-assisted SIR, respectively. Further extensions to expectile-assisted SAVE and
expectile-assisted directional regression are described in
Section 4. Some practical issues such as tuning parameter selection are discussed in Section 5. Extensive simulation studies are provided in Section 6 and we conclude the paper with a real data analysis in Section 7. All proofs and additional simulation results are relegated to the Appendix.

\section{Population level development of expectile-assisted SIR}
Expectiles were first introduced by Newey and Powell (1987) in the seminal asymmetric least squares paper.
It has gained popularity in finance and risk management for estimating the expected shortfall and value at risk. See, for example, Kim and Lee (2016), Daouia, Girard and Stupfler (2018), and Chen (2018).
For  $0<\uptau<1$,  denote $f_{\uptau}(\bm{X})$ as the $\uptau$-th expectile of the conditional distribution of $Y$ given $\bm{X}$. Then
\begin{align}
  f_{\uptau}(\bm{x}) &= \argmin_a \ E\big\{\phi_\uptau(Y - a) | \bm{X}=\bm{x} \big \},
   \label{als_loss}
\end{align}  where $\phi_\uptau(\cdot)$ is known as the asymmetric loss function and is defined as
\begin{align}\phi_\uptau(c) = \begin{cases}
 (1-\uptau)c^2, \  \ \mbox{if }c \leq 0,\\
 \uptau c^2, \  \ \mbox{if }c > 0.
\end{cases} 
   \label{als_loss2}
\end{align}

\begin{proposition}\label{proposition1}
 For  $0<\uptau_1 < \cdots < \uptau_k<1$, let
 $\bm{\xi}_{\bm{X}} = \big (f_{\uptau_1}(\bm{X}), \ldots, f_{\uptau_k}(\bm{X}) \big )^\top$. Then
 $\mathcal{S}_{\bm{\xi_X} | \bm{X}}\subseteq \mathcal{S}_{Y|\bm{X}}$.
\end{proposition}

\noindent Proposition \ref{proposition1} suggests that we can recover the central space $\mathcal{S}_{Y|\bm{X}}$ through estimation of the central space for the regression of $\bm{\xi_X}$ on $\bm{X}$. We implicitly assume that $f_{\uptau_\ell}(\bm{X})$ from (\ref{als_loss}) is well-defined for $\ell=1,\ldots,k$.

For $\bm{\xi_X} \in \mathbb{R}^k$, the original SIR can not be applied directly due to the multivariate response. Let $\bm{T} \in \mathbb{R}^k$ be a random vector. We follow Li, Wen, and Zhu (2008) and apply SIR for the regression between $ \bm{\xi_X}^\top \bm{T}$ and $\bm{X}$ instead.
Let $E(\bm{X})=\bm{\mu}$ and ${\rm Var}(\bm{X})=\bm{\Sigma}$. Then $\bm{Z} = \bm{\Sigma}^{-1/2}(\bm{X}-\bm{\mu})$ denotes the standardized predictor. 
Let $J_1(\bm{T}),\ldots, J_H(\bm{T})$ be the partition of the support of $\bm{\xi_X}^\top \bm{T}$. For $h=1,\ldots, H$, denote $I_h(\bm{T})$ as the indicator function of
$\bm{\xi_X}^\top \bm{T}\in J_h(\bm{T})$.
Define
\begin{align}
     \bm{M}(\bm{T}) = \sum_{h=1}^H p_h(\bm{T}) \bm{\mu}_h(\bm{T}) \bm{\mu}_h^\top(\bm{T}),  
    \label{EASIR}
\end{align}
 where   
    $p_h(\bm{T}) = E\{I_h(\bm{T}) \}$ and $ \bm{\mu}_h(\bm{T})  = E\{\bm{Z} | \bm{\xi_X}^\top \bm{T} \in J_h(\bm{T})\}$.

Before we state the next result, we need the following  linear conditional mean (LCM) assumption, which is a common assumption in the sufficient dimension reduction literature.  

\begin{assumption}\label{linearity}
$E(\bm{X} | \bm{B}^\top{\bm{X}})$ is a linear function of $\bm{B}^\top{\bm{X}}$, where  $\bm{B}$ is a basis of $\mathcal{S}_{Y|\bm{X}}$.
\end{assumption}

\begin{proposition}\label{projectiveSIR}
Let $\bm{T}$ be a random vector uniformly distributed on the unit sphere $\mathbb{S}^k$. Then under Assumption \ref{linearity}, ${\rm span}\big( \bm{\Sigma}^{-1/2} E\{ \bm{M}(\bm{T}) \} \big )  \subseteq  \mathcal{S}_{Y|\bm{X}}$.
\end{proposition}

\noindent Here ${\rm span(\cdot)}$ denotes the column space, and  the expectation $E\{ \bm{M}(\bm{T}) \} $ is over the distribution of $\bm{T}$.
We remark that the LCM assumption is not needed in Proposition \ref{proposition1}, and it is only needed in Proposition \ref{projectiveSIR} because the classical SIR requires the LCM assumption.

\section{Sample level algorithm of expectile-assisted SIR}


\subsection{ A toy example}
 
\begin{figure}[t]
	\centering
	\includegraphics[width=5.5in,]{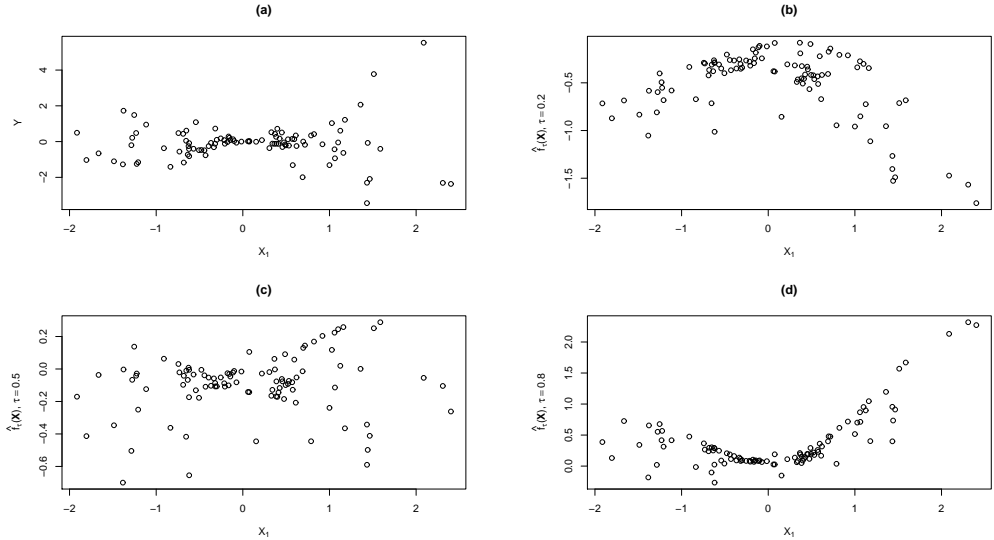}
	\caption{A toy example with $Y = X_1\epsilon$. (a) $Y$ versus $X_1$; (b) $\hat{f}_{\uptau}(\bm{X})$ versus $X_1$ with $\uptau=0.2$; (c) $\hat{f}_{\uptau}(\bm{X})$ versus $X_1$ with $\uptau=0.5$; (d) $\hat{f}_{\uptau}(\bm{X})$ versus $X_1$ with $\uptau=0.8$.}
	\label{toy} 
	\end{figure}
 
 We first consider a toy example to fix the idea of expectile-assisted sufficient dimension reduction. Let $\bm{X}=(X_1,X_2)^\top$, where $X_1\sim N(0,1)$, $X_2\sim N(0,1)$, and $X_1$ is independent of $X_2$. Let $Y = X_1\epsilon$, where $\epsilon\sim N(0,1)$ and $\epsilon$ is independent of $\bm{X}$. Given
 $100$ data points generated from this model, the scatter plot of $Y$ versus $X_1$ is provided in panel (a) of Figure \ref{toy}, which shows a clear heteroscedastic trend. While the classical sufficient dimension reduction problem aims to recover the central space $\mathcal{S}_{Y|\bm{X}}$ through the original response $Y$, Proposition \ref{proposition1} indicates that we could consider the alternative dimension reduction problem with the expectile $f_{\uptau}(\bm{X})$ as the new response. In practice,  $f_{\uptau}(\bm{X})$ has to be replaced by its sample level estimator  $\hat{f}_{\uptau}(\bm{X})$, which will be examined in Section 3.2.   In panels (b), (c) and (d), the scatter plot of $\hat{f}_{\uptau}(\bm{X})$ versus $X_1$ is shown for $\uptau=0.2$, $0.5$ and $0.8$, respectively. It is obvious from the scatter plots that  $\hat{f}_{\uptau}(\bm{X})$  depends on $X_1$, and the strength of such dependence may vary when we change the expectile level $\uptau$.

\subsection{Kernel expectile regression}

Given an i.i.d. sample
$\{(\bm{X}_i,Y_i): i=1,\ldots, n\}$, we explain how to estimate the $\uptau$-th expectile $f_{\uptau}(\bm{X})$ of the conditional distribution of $Y$ given $\bm{X}$ in this section. This step is the same for expectile-assisted SIR
and the other expectile-assisted  inverse regression methods to be discussed in Section 4.
The original estimator in Newey and Powell (1987) focused on expectiles in linear regression. To estimate the conditional expectiles in nonlinear models, Yao and Tong (1996) proposed a local linear polynomial estimator. More recently, Yang, Zhang and Zou (2018) developed a  reproducing kernel Hilbert space (RKHS) estimator for flexible expectile regression. We adapt the RKHS estimator with the following Gaussian radial basis kernel 
\begin{align}
     K(\bm{X}_i,\bm{X}_j) = \exp(-r\|\bm{X}_i-\bm{X}_j \|^2),
    \label{rbf}
\end{align}
where $r$ is a tuning parameter and $\|\cdot\|$ denotes the Euclidean norm. Let $\mathbb{H}_K$ be the RKHS generated from the kernel function (\ref{rbf}).
As an element of $\mathbb{H}_K$,  $f_{\uptau}(\bm{X})$ evaluated at $\bm{X}=\bm{X}_i$  can be estimated by
\begin{align} \label{f hat}
   \hat{f}_\uptau(\bm{X}_i) = \hat{\alpha}_{0,\uptau} + \displaystyle\sum _{j=1}^n \hat{\alpha}_{j,\uptau} K(\bm{X}_i,\bm{X}_j).
\end{align}
Let $\bm{\hat{\alpha}}_\uptau = (\hat{\alpha}_{0,\uptau},\hat{\alpha}_{1,\uptau},\ldots,\hat{\alpha}_{n,\uptau})$  and $\bm{\alpha}_\uptau =(\alpha_{0,\uptau},\alpha_{1,\uptau},\ldots,\alpha_{n,\uptau})$.
Then $\bm{\hat{\alpha}}_\uptau$ in (\ref{f hat}) is the minimizer of the regularized empirical risk function on  $\mathbb{H}_K$
\begin{align}
\begin{split}
\bm{\hat{\alpha}}_\uptau =     \underset{\bm{\alpha}_\uptau}{\operatorname{argmin}} & \hspace{0.11in} \displaystyle\sum_{i=1}^n \phi_\uptau \Big(Y_i -\alpha_{0,\uptau} -  \displaystyle\sum _{j=1}^n \alpha_{j,\uptau} K(\bm{X}_i,\bm{X}_j)\Big)\\
& + \lambda \displaystyle\sum _{i=1}^n \displaystyle\sum _{j=1}^n \alpha_{i,\uptau}\alpha_{j,\uptau} K(\bm{X}_i,\bm{X}_j),
\end{split}
\label{obj1}
\end{align}
where $\phi_\uptau(\cdot)$ is defined in (\ref{als_loss}) and $\lambda$ is a tuning parameter. The optimization (\ref{obj1}) and the evaluation (\ref{f hat}) can be done very efficiently in the ${\rm KERE}$ package in ${\rm R}$. The choices for the tuning parameters $r$ in  (\ref{rbf}) and $\lambda$ in (\ref{obj1}) are discussed in Section 5.

\subsection{Projective resampling for multiple response SIR}

Given an i.i.d. sample
$\{(\bm{X}_i,Y_i): i=1,\ldots, n\}$, the sample level expectile-assisted SIR algorithm is as follows. 


\begin{enumerate}
     \item For a given integer $k$, specify $0 < \uptau_1 <\cdots<\uptau_k < 1$. For $i=1,\ldots,n$, calculate 
     $\hat{\bm{\xi}}_{\bm{X}_i} = \big (\hat f_{\uptau_1}(\bm{X}_i), \cdots, \hat f_{\uptau_k}(\bm{X}_i) \big )^\top$, where the $\ell$-th component of 
     $\hat{\bm{\xi}}_{\bm{X}_i}$ is given by  (\ref{f hat}) with $\uptau=\uptau_\ell$. 
     \item Let $\hat{\bm{\mu}}=n^{-1}\sum_{i=1}^n \bm{X}_i$ and $\hat{\bm{\Sigma}}=n^{-1}\sum_{i=1}^n ( \bm{X}_i-\hat{\bm{\mu}})( \bm{X}_i-\hat{\bm{\mu}})^\top$. Calculate standardized predictors $\bm{\hat{Z}}_i = \hat{\bm{\Sigma}}^{-1/2}(\bm{X}_i-\hat{\bm{\mu}})$ for $i=1,\ldots,n$.
   \item For a given integer $N$, generate an i.i.d. sample $\bm{t}^{(1)},\ldots,\bm{t}^{(N)}$ from the uniform distribution on the unit sphere $\mathbb{S}^k$.
   For $j=1,\ldots,N$, let $J_1(\bm{t}^{(j)}), \ldots$, $J_H(\bm{t}^{(j)})$ be the partition of the support of $\hat{\bm{\xi}}_{\bm{X}}^\top \bm{t}^{(j)}$. For $h=1,\ldots, H$, denote $I_{hi}(\bm{t}^{(j)})$ as the indicator function of
$\hat{\bm{\xi}}_{\bm{X}_i}^\top \bm{t}^{(j)}\in J_h(\bm{t}^{(j)})$.
   \item We now calculate the sample version of $E\{ \bm{M}(\bm{T}) \}$.
   
   \begin{enumerate}
\item[4.1] For $j=1,\ldots,N$, let $\hat p_h(\bm{t}^{(j)}) = n^{-1}\sum_{i=1}^n I_{hi}(\bm{t}^{(j)})$ and
$\hat{\bm{\mu}}_h(\bm{t}^{(j)})= \{n\hat p_h(\bm{t}^{(j)}) \}^{-1}\sum_{i=1}^n \bm{\hat{Z}}_i I_{hi}(\bm{t}^{(j)})$. Then the sample estimator of (\ref{EASIR}) with $\bm{T}=\bm{t}^{(j)}$
becomes
$$\hat{\bm{M}}(\bm{t}^{(j)})= \sum_{h=1}^H \hat p_h(\bm{t}^{(j)})\hat{\bm{\mu}}_h(\bm{t}^{(j)})  \hat{\bm{\mu}}_h^\top(\bm{t}^{(j)}).$$
\item[4.2] Calculate $\hat{\bm{M}}(\bm{T})=N^{-1}\sum_{j=1}^N \hat{\bm{M}}(\bm{t}^{(j)})$. 
\end{enumerate}

\item For  a given structural dimension $d$, let $(\hat{\bm{v}}_1,..,\hat{\bm{v}}_d)$ be the eigenvectors corresponding to the $d$ leading eigenvalues of
$\hat{\bm{M}}(\bm{T})$. The final estimator of $\mathcal{S}_{Y|\bm{X}}$ is then ${\rm span}(\hat{\bm{B}})$, where $\hat{\bm{B}}=(\hat{\bm{\Sigma}}^{-1/2}\hat{\bm{v}}_1,..,
\hat{\bm{\Sigma}}^{-1/2}\hat{\bm{v}}_d)$.
\end{enumerate}

\noindent In the numerical studies in Section 6 and the real data analysis in Section 7, we fix $N=1000$, $k=9$, and set $\uptau_\ell=10^{-1}\ell$ for $\ell=1,\ldots,9$. The effects of different $N$ and $k$ are studied in the Appendix. Our experience suggests that the proposed method is not very sensitive to the choice of $N$ and $k$.

\section{Extensions of SAVE and directional regression}

Expectile-assisted dimension reduction is a very general framework, and can be readily generalized to other moment-based methods such as SAVE and directional regression. We focus on the population level development of expectile-assisted SAVE and expectile-assisted directional regression in this section.

Recall that $I_h(\bm{T})$ denotes the indicator function of
$\bm{\xi_X}^\top \bm{T}\in J_h(\bm{T})$, $p_h(\bm{T}) = E\{I_h(\bm{T}) \}$, and $ \bm{\mu}_h(\bm{T})  = E\{\bm{Z} | \bm{\xi_X}^\top \bm{T} \in J_h(\bm{T})\}$.
Define 
$$\bm{G}(\bm{T})=\sum_{h=1}^H p_h(\bm{T})\{\bm{V}_h(\bm{T})-  \bm{\mu}_h(\bm{T}) \bm{\mu}_h^\top(\bm{T})  \}^2,$$
where $\bm{V}_h(\bm{T})=E\{\bm{Z}\bm{Z}^\top - \bm{I}_p | \bm{\xi_X}^\top \bm{T} \in J_h(\bm{T})\}$. In addition to the LCM assumption, we need the constant conditional variance (CCV) assumption as follows 
\begin{assumption}\label{ccv}
${\rm Var}(\bm{X}| \bm{B}^\top{\bm{X}})$ is a nonrandom matrix, where  $\bm{B}$ is a basis of $\mathcal{S}_{Y|\bm{X}}$.
\end{assumption}


\begin{proposition}\label{projectiveSAVE}
Let $\bm{T}$ be a random vector uniformly distributed on the unit sphere $\mathbb{S}^k$. Then under Assumptions \ref{linearity} and \ref{ccv}, ${\rm span}\big( \bm{\Sigma}^{-1/2} E\{ \bm{G}(\bm{T}) \} \big )  \subseteq  \mathcal{S}_{Y|\bm{X}}$.
\end{proposition}

In a similar fashion,
define
\begin{align*}
    \bm{F}(\bm{T}) &=  2\sum_{h=1}^H p_h(\bm{T}) \bm{V}_h(\bm{T})\bm{V}_h(\bm{T}) + 2\left \{\sum_{h=1}^H p_h(\bm{T}) \bm{\mu}_h(\bm{T}) \bm{\mu}_h^\top(\bm{T})  \right \}^2 \\
    &+ 2\left\{ \sum_{h=1}^H p_h(\bm{T}) \bm{\mu}_h^\top(\bm{T}) \bm{\mu}_h(\bm{T}) \right\} \left\{\sum_{h=1}^H p_h(\bm{T}) \bm{\mu}_h(\bm{T}) \bm{\mu}_h^\top(\bm{T})  \right \},
\end{align*}
and we have
\begin{proposition}\label{projectivedr}
Let $\bm{T}$ be a random vector uniformly distributed on the unit sphere $\mathbb{S}^k$. Then under Assumptions \ref{linearity} and \ref{ccv}, ${\rm span}\big( \bm{\Sigma}^{-1/2} E\{ \bm{F}(\bm{T}) \} \big )  \subseteq  \mathcal{S}_{Y|\bm{X}}$.
\end{proposition}

\noindent Based on Proposition \ref{projectiveSAVE} and Proposition \ref{projectivedr}, we may update step 4 and step 5 of the  expectile-assisted  SIR algorithm to get 
the sample estimators of 
$\bm{\Sigma}^{-1/2}  E\{ \bm{G}(\bm{T}) \} $ and $\bm{\Sigma}^{-1/2}  E\{ \bm{F}(\bm{T}) \} $. We refer to them as the expectile-assisted SAVE estimator and the expectile-assisted directional regression estimator, respectively.

\section{Additional issues}

\subsection{Selecting tuning parameters}

We first  discuss the choice of $r$ in  (\ref{rbf}). For the Gaussian radial basis kernel, Li, Artemiou and Li (2011) suggested using
\begin{align}\label{r value}
    r = 1/\gamma^2 \mbox{ with } \gamma = \frac{2}{n(n-1)} \sum_{i=1}^{n-1}\sum_{j=i+1}^n \|\bm{X}_i - \bm{X}_j\|,
\end{align}
where $(\bm{X}_1,\ldots,\bm{X}_n)$ is an i.i.d. sample. The effect of different choices of $r$ is examined in the Appendix, and it is shown that the proposed methods are not sensitive to the choice of $r$ when it  varies around $1/\gamma^2$.

Now we turn our attention to selecting $\lambda$ in (\ref{obj1}). 
For a given $\lambda$, denote the final estimator from the algorithm in Section 3.3  as 
$\hat{\bm{B}}_\lambda$. 
Distance correlation (Sz{\'e}kely, Rizzo  and  Bakirov, 2007) is a good measure of linear as well as nonlinear dependence. Denote 
$\rho^2(Y, \bm{B}^\top \bm{X})$ as the squared distance correlation between $Y$ and $\bm{B}^\top \bm{X}$. It is known that $\rho^2(Y, \bm{B}^\top \bm{X})=0$ if and only if $Y$ is independent of $\bm{B}^\top \bm{X}$.
Given an i.i.d. sample $\{(\bm{X}_i,Y_i): i=1,\ldots, n\}$ and $\hat{\bm{B}}_\lambda$, the  sample  squared distance correlation between  $Y$ and $\hat{\bm{B}}_\lambda^\top \bm{X}$ is denoted as $\hat \rho^2(Y, \hat{\bm{B}}_\lambda^\top \bm{X})$. From a set of candidate values for $\lambda$, we choose $\lambda$ such that the sample squared distance correlation  $\hat \rho^2(Y, \hat{\bm{B}}_\lambda^\top \bm{X})$ is maximized. The comparison between the data-driven approach and using a prespecified $\lambda$ is provided in  the Appendix, which indicates that the data-driven approach works well in practice.

SIR, SAVE, directional regression, and their expectile-assisted counterparts all rely on slicing the continuous response. For the number of slices $H$, existing sufficient dimension reduction literature suggests that SIR is not very sensitive to $H$ as it only involves intraslice means (Zhu, L. X. and Ng, K. W., 1995). SAVE, on the other hand, involves intraslice variances and is more sensitive to $H$ than SIR (Li, Y. and Zhu, L. X., 2007). We examine the effect of $H$ in the Appendix. It seems that expectile-assisted SIR and expectile-assisted directional regression are not very sensitive to the choice of $H$, while expectile-assisted SAVE prefers smaller values of $H$.

In step 5 of the sample level algorithm in Section 3.3, we need the structural dimension $d$, which has to be estimated in practice. The problem of estimating the structural dimension is known as order determination. Sequential test is a common method for order determination in the sufficient dimension reduction literature. For order determination based on asymptotic sequential test, one may refer to Chapter 9 of Li (2018). Cook and Yin (2001) proposed a permutation sequential test approach for order determination, which can be directly applied to our proposed expectile-assisted methods. The comparison between  the asymptotic sequential test and the permutation sequential test is provided in the Appendix.


\subsection{Pooled marginal estimators}

In Cook and Setodji (2003), Saracco (2005),  Yin and Bura (2006), Barreda, Gannoun and Saracco (2007), Coudret, Girard and Saracco  (2014), pooled marginal estimators are proposed for sufficient dimension reduction with multiple responses. 
Without loss of generality, we propose the pooled marginal expectile-assisted SIR in this section. The extensions to SAVE and directional regression are similar and thus omitted. 

Recall that for $\ell=1,\ldots, k$, $f_{\uptau_\ell}(\bm{X})$ denotes the $\uptau_\ell$-th conditional expectile of $Y$ given $\bm{X}$. Let $J_{1,\ell},\ldots,J_{H,\ell}$ be a partition for the support of $f_{\uptau_\ell}(\bm{X})$. Let 
$p_{h,\ell} = E\{f_{\uptau_\ell}(\bm{X})\in J_{h,\ell}\}$, $ \bm{\mu}_{h,\ell}  = E\{\bm{Z} | f_{\uptau_\ell}(\bm{X})\in J_{h,\ell}\}$, and
$\bm{M}_\ell = \sum_{h=1}^H p_{h,\ell} \bm{\mu}_{h,\ell}  \bm{\mu}_{h,\ell} ^\top$. Define $\widetilde{\bm{M}}=(\bm{M}_1,\ldots,\bm{M}_k)$, and we have 

\begin{proposition}\label{msir}
Under Assumption \ref{linearity}, ${\rm span}\big( \bm{\Sigma}^{-1/2} \widetilde{\bm{M}}\big )  \subseteq  \mathcal{S}_{Y|\bm{X}}$.
\end{proposition}

\noindent The marginal approach essentially considers $k$ univariate response sufficient dimension reduction problems separately and then assemble the individual estimators for each response to get the final estimator. At the sample level, 
the estimator of the central space consists of  the left singular vectors of the sample version of  $\bm{\Sigma}^{-1/2} \widetilde{\bm{M}}$. We refer to it as the pooled marginal expectile-assisted SIR estimator. 


\section{Simulation studies}\label{Sim_study}

We examine the empirical performances of  our proposals through synthetic examples in this section. 
The predictor $\bm{X}$ is generated from $N(\bm{0},\bm{I}_p)$ with $p=6$ or $p=20$. The first six components of $\bm{\beta}_1\in  \mathbb{R}^p$ is $(1,1,1,0,0,0)$, and the first six  components of $\bm{\beta}_2\in  \mathbb{R}^p$ is $(1,0,0,0,1,3)$. The remaining components of $\bm{\beta}_1$ and $\bm{\beta}_2$ are all zero when $p=20$. The response $Y$ is generated as follows:
\begin{enumerate}
\item[I] : $Y = 0.4(\bm{\beta}_1^\top\bm{X})^2 + 3\sin(\bm{\beta}_2^\top\bm{X}/4) + \sigma\epsilon$,
\item[II] : $Y = 3\sin(\bm{\beta}_1^\top\bm{X}/4) + 3\sin(\bm{\beta}_2^\top\bm{X}/4) + \sigma\epsilon$,
\item[III] : $Y = 0.4(\bm{\beta}_1^\top\bm{X})^2 + |\bm{\beta}_2^\top\bm{X}|^{1/2} + \sigma\epsilon$,
\item[IV] : $Y = 3\sin(\bm{\beta}_2^\top\bm{X}/4) + \{1+(\bm{\beta}_1^\top\bm{X})^2\}\sigma\epsilon$,
\item[V]: $Y = \bm{\beta}_1^\top\bm{X}\epsilon$,
\end{enumerate}
where $\sigma = 0.2$, $\epsilon \sim N(0,1)$, and $\epsilon$ is independent of $\bm{X}$. 
Models I through IV are the same models used in Li and Wang (2007), and model V is similar to the toy example in Section 3.1. Following Li and Wang (2007), two sample size settings are considered. For $n=100$, we set $p=6$ and number of slices $H=5$. For $n=500$, we set $p=20$ and $H=10$.

We compare SIR, SAVE,  directional regression (DR), expectile-assisted SIR (EA-SIR), expectile-assisted SAVE (EA-SAVE), and expectile-assisted directional regression (EA-DR).  
Quantile-slicing mean estimation (QUME) and quantile-slicing variance estimation (QUVE) (Kim, Wu and Shin, 2019) are also included for the comparison. For models I through IV, the basis for the central space is $\bm{B}=(\bm{\beta}_1,\bm{\beta}_2)$, and the central space basis for model V is $\bm{B}=\bm{\beta}_1$. For estimator $\hat{\bm{B}}$, we measure its performance by 
$\Delta=\|\bm{P}_{\bm{B}}-\bm{P}_{\hat{\bm{B}}}\|_F$.
Here $\bm{P}_{\bm{B}} = \bm{B}(\bm{B}^\top \bm{B})^{-1}\bm{B}^\top$,  $\bm{P}_{\hat{\bm{B}}} = \hat{\bm{B}}(\hat{\bm{B}}^\top \hat{\bm{B}})^{-1}\hat{\bm{B}}^\top$, and $\|\cdot\|_F$ denotes the matrix   Frobenius norm. 
Smaller values of $\Delta$ mean better performances.
We fix the number of projections as $N=1000$ and the number of expectile levels as $k=9$ for all three expectile-assisted methods. Furthermore, $r$ for the Gaussian radial basis kernel is set as in (\ref{r value}), $\lambda$ for the regularization term in (\ref{obj1}) is chosen in a data-driven manner as described in  Section 5.1. More simulation studies are provided in the Appendix for different choices of $H$, $N$, $k$, $r$ and $\lambda$.



\begin{table}[t]
	\scalebox{0.85}{ 
		\begin{tabular}{c|ccc|ccc|cc}
			\hline
			Model & SIR  & EA-SIR  & QUME  & SAVE  & EA-SAVE & QUVE & DR   & EA-DR    \\
			\hline
			
			\multirow{2}{*}{I}
			& 1.648	& 1.343	&  1.512  & 0.626  & 0.554 & 0.939 		&	 0.384	& 0.345	\\
			& (0.043)& (0.058)	& (0.047)& (0.059)	& (0.050)	& (0.052)	& (0.041)	& (0.029) \\ \hline
			
			\multirow{2}{*}{II}
			& 1.521	& 1.567	& 1.518	& 1.565	& 1.543	& 1.737 & 1.492	& 1.497	 \\
			& (0.046)	& (0.046)	&(0.049)	& (0.047)	& (0.048)	& (0.040)	& (0.051)	&(0.050) \\ \hline
			
			\multirow{2}{*}{III} 
			& 2.620	& 2.308	& 2.722	& 0.652 & 0.547	& 0.287 & 0.638	 & 0.543		 \\
			& (0.063)	& (0.060)	& (0.067)	& (0.050)	& (0.046)	& (0.034)	& (0.049)	& (0.048)  \\ \hline
			
			\multirow{2}{*}{IV}
			& 1.700	& 1.396 & 1.556	& 1.598	& 1.247 	& 1.734 		& 1.557	& 1.177 \\
			& (0.034)& (0.054)	& (0.047)	& (0.046)	& (0.056)	& (0.047)	& (0.046)	& (0.056) \\ \hline
			
			\multirow{2}{*}{V}
			& 1.667 & 1.484 & 1.513		& 0.572 & 0.792		& 0.967 	& 0.561		& 0.799\\
			& (0.037)	& (0.052)	& (0.048)	& (0.046)	& (0.061)	& (0.059)	& (0.045)	& (0.064) \\ \hline
			
		\end{tabular}
	}
	\caption{Results based on $(n,p)=(100,6)$. The average of $\Delta$ and its standard error (in parentheses) are reported based on 100 repetitions.}
	\label{case1}
\end{table}

For the  $(n,p)=(100,6)$ setting, we summarize the simulation results based on $100$ repetitions in Table  \ref{case1}. 
First we compare SIR with EA-SIR and QUME, which all belong to first-order inverse regression estimators. We see that while QUME improves over SIR with the exception of model III, EA-SIR has the best overall performance among these three methods. Next we compare SAVE with EA-SAVE and QUVE, which are all second-order inverse regression methods. EA-SAVE again has the  best overall performance in this group. It is interesting to see that QUVE is worse than both SAVE and EA-SAVE for all models other than model III, where the link functions are symmetric.  When we compare SIR-based methods with SAVE-based methods, we see that SIR-based methods do not work as well for models I, III, IV, and V.  This is due to the fact that at least one link function in the mean component is symmetric for models I, III and IV, and SIR is known to be ineffective in the presence of symmetric link functions. EA-SIR and QUME inherit this limitation. Although the error term in model V involves an asymmetric linear function of $\bm{X}$, we have seen in panel (a) of Figure \ref{toy} that there is still symmetry about the $y$-axis in this type of heteroscedastic model. We note that EA-SIR improves over SIR in both model IV and V with heteroscedastic error. DR is very competitive across all five models as it is not sensitive to the shape of the link functions. EA-DR further improves over DR in three out of five models and enjoys the best overall performance.

The simulation results for the $(n,p)=(500,20)$ setting are summarized in Table \ref{case2}. We observe similar results as in Table \ref{case1}: the overall performance of EA-SIR is better than SIR and QUME; the overall performance of EA-SAVE is better than SAVE and QUVE; DR and EA-DR enjoy the best overall performances. Furthermore, SAVE-based methods are significantly worse than their SIR-based counterparts for model II, where both link functions are monotone. It is known in the sufficient dimension reduction literature that SAVE may not be very efficient with monotone link functions (Li and Wang, 2007). EA-SAVE and  QUVE also have this limitation.

\begin{table}[t]
	\scalebox{.85}{ 
		\begin{tabular}{c|ccc|ccc|cc}
			\hline
			Model & SIR  & EA-SIR & QUME & SAVE   & EA-SAVE & QUVE  & DR   & EA-DR     \\ \hline
			
			\multirow{2}{*}{I}
			& 1.845	& 1.707 & 1.724 & 1.114 	& 0.454 & 0.846 	& 0.245		& 0.265		\\
			& (0.026)	& (0.037)	& (0.035)	& (0.061)	& (0.017)	& (0.040)	& (0.007)	& (0.009) \\ \hline
						
			\multirow{2}{*}{II}			
			& 1.564 & 1.685 & 1.871	& 1.796 & 1.845 & 2.017 	& 1.710			& 1.735		\\
			& (0.036)	& (0.033)	& (0.016)	& (0.023)	& (0.019)	& (0.011)	& (0.029)	& (0.032) \\ \hline
						
			\multirow{2}{*}{III}
			& 3.594	& 3.447 & 3.534  & 0.451 & 0.364 & 0.710	& 0.443	& 0.355		 \\
			& (0.028)& (0.036)& (0.031)& (0.016)	& (0.012)	& (0.050)	& (0.015)	& (0.013) \\ \hline
						
			\multirow{2}{*}{IV}
			& 1.908 & 1.832 & 1.904	& 1.747 & 1.524 & 2.053	& 1.584			& 1.473		 \\
			& (0.016)	& (0.027)	& (0.015)	& (0.040)	& (0.042)	& (0.013)	& (0.041)	& (0.045) \\ \hline	
			
			\multirow{2}{*}{V}
			& 1.915 & 1.850& 1.842			& 0.283  & 0.388 	& 0.313 	& 0.280		& 0.373\\
			& (0.011)	& (0.019)	& (0.018)	& (0.011)	& (0.044)	& (0.022)	& (0.011)	& (0.043) \\ \hline
			
		\end{tabular}
	}
	\caption{Results based on $(n,p)=(500,20)$. The average of $\Delta$ and its standard error (in parentheses) are reported based on 100 repetitions.}
	\label{case2}
\end{table}

Next we compare our proposed expectile-assisted methods based on random projections with the pooled marginal estimators described in Section 5.2.  We denote the pooled marginal expectile-assisted SIR as mEA-SIR. Similarly, mEA-SAVE and mEA-DR denote the corresponding pooled marginal estimators for SAVE and DR.
For this comparison, we fix $p=6$, $H=5$, and consider $n=50$, $100$ and $150$. The results based on $100$ repetitions are summarized in Table \ref{pm}. The pooled marginal estimators have decent performances, which confirms the result of Proposition \ref{msir}. As sample size increases, the performances of all projective resampling methods as well as the pooled marginal methods improve. The pooled marginal estimators are outperformed by the corresponding projective resampling estimators in $5$ out of $9$ cases for model I, and  in $7$ out of $9$ cases for model II. This confirms the finding in Li, Wen and Zhu (2008) that projective resampling is more efficient than the pooled marginal estimators.

\begin{table}[t]
	\scalebox{.85}{ 
\begin{tabular}{c|c|cc|cc|cc}
\hline
Model   & $n$   & EA-SIR & mEA-SIR  & EA-SAVE  & mEA-SAVE & EA-DR  & mEA-DR  \\
\hline

{\multirow{6}{*}{I}}

&  {\multirow{2}{*}{50}}  & 1.531 & 1.481 & 1.772 & 2.050 & 0.893 & 0.934 \\
& & (0.057) & (0.059) & (0.072) & (0.065) & (0.057) & (0.059) \\
\cline{2-8}  

&  {\multirow{2}{*}{100}}  & 1.343	& 1.302	& 0.554	& 0.678	& 0.345	& 0.368 \\
& & (0.058) & (0.059) & (0.050) & (0.061) & (0.029) & (0.032) \\
\cline{2-8}  

&  {\multirow{2}{*}{150}}  & 1.274	& 1.160	& 0.270	& 0.270	& 0.174	& 0.190 \\
& & (0.062) & (0.057) & (0.033) & (0.033) & (0.012) & (0.015) \\
\hline

{\multirow{6}{*}{II}}

&  {\multirow{2}{*}{50}}  & 1.540 & 1.558 & 1.698 & 2.144 & 1.574 & 1.571 \\
& & (0.050) & (0.046) & (0.048) & (0.071) & (0.045) & (0.044) \\
\cline{2-8}  

&  {\multirow{2}{*}{100}}  & 1.567	& 1.564	& 1.543	& 1.612	& 1.497	& 1.521 \\
& & (0.046) & (0.043) & (0.048) & (0.043) & (0.050) & (0.049) \\
\cline{2-8}  

&  {\multirow{2}{*}{150}}  & 1.406	& 1.473	& 1.426	& 1.474	& 1.425	& 1.460 \\
& & (0.055) & (0.052) & (0.048) & (0.050) & (0.052) & (0.051) \\
\hline

\end{tabular}}
\caption{Results based on $p=6$. The average of $\Delta$ and its standard error (in parentheses) are reported based on 100 repetitions.}
\label{pm}
\end{table} 

%
%
%
%
%
%
%
%
%

\section{Analysis of the Big Mac data}

The Big Mac data contains 10 economic variables from 45 cities around the world in 1991. The data  can be downloaded at 
\url{http://www.stat.umn.edu/RegGraph/data/Big-Mac.lsp}. The response $Y$ is the minutes of labor needed to buy a Big Mac. The detailed description of the predictors can be found at the above website. Following the discussions in Li (2008) (page 92), we apply the optimal Box-Cox transformation (Box and Cox, 1964) before we compare different dimension reduction methods. Denote the predictors after the Box-Cox transformation   as $\bm{X}=(X_1, \ldots, X_9)^\top$. As suggested in Li (2018) (page 139, Table 9.1), we use structural dimension $d=1$ for this data.

First, we use leave-one-out cross validation to compare the performances of six methods: SIR, DR, EA-SIR, EA-DR, QUME and QUVE. Consider the training set to be the 44  observations after removing one observation from the entire data. For a given dimension reduction method,
denote $\hat{\bm{\beta}}^{(-i)}\in \mathbb{R}^9$ as the central space estimator based on the $i$-th training set, where the $i$-th observation $(\bm{X}_{i}, Y_i)$ is removed, $i=1,\ldots, 45$.
For a fixed expectile level $\uptau$, we fit the kernel expectile regression between $Y$ and $\bm{X}^\top\hat{\bm{\beta}}^{(-i)} $ for the $i$-th training set, and denote the estimated $\uptau$-th conditional expectile function as $\hat f_\uptau^{(-i)}$. Then we evaluate $\hat f_\uptau^{(-i)}$ at $\bm{X}_{i}^\top\hat{\bm{\beta}}^{(-i)}$ and compare it to $Y_i$ through the asymmetric least squares loss function $\phi_\uptau\{Y_i-\hat f_\uptau^{(-i)}(\bm{X}_{i}^\top\hat{\bm{\beta}}^{(-i)})\}$, where $\phi_\uptau$ is defined in (\ref{als_loss2}). Repeat this process for all $i$, and the average asymmetric least squares loss is defined as 
 \begin{align*}
 \delta_\uptau=\cfrac{1}{45} \displaystyle\sum_{i=1}^{45}\phi_\uptau\{Y_i-\hat f_\uptau^{(-i)}(\bm{X}_{i}^\top\hat{\bm{\beta}}^{(-i)})\}.
 \end{align*} 
 For SIR, DR, EA-SIR and EA-DR, we use $H=2$ or $H=4$ to estimate $\hat{\bm{\beta}}^{(-i)}$. Fix $\uptau$ to be $0.2$, $0.5$ and $0.8$, and the $\delta_\uptau$ values are summarized in Table \ref{bigmac}. We see that the expectile-assisted methods improve over their classical counterparts. For each fixed $\tau$, quantile-based methods have the largest $\delta_\uptau$ values, and EA-SIR with $2$ slices consistently has the smallest  $\delta_\uptau$.

\begin{table}[t] 
	\scalebox{0.85}{ 
	\begin{tabular}{c|c|c|c|c|c|c|c|c|c|c}
	\hline
\multirow{2}{*}{$\uptau$}& \multicolumn{2}{|c|}{SIR} & \multicolumn{2}{|c|}{EA-SIR}
& \multicolumn{2}{|c|}{DR} & \multicolumn{2}{|c|}{EA-DR} & \multirow{2}{*}{QUME} & \multirow{2}{*}{QUVE}\\
\cline{2-9}
& $H=2$ & $H=4$ & $H=2$ & $H=4$ & $H=2$ & $H=4$ & $H=2$ & $H=4$ & &\\
\hline
$0.2$ & 1165 & 831 & 615 & 711  & 1294 & 1330 & 1028 & 951 & 2009 & 1332\\
	\hline
$0.5$ & 858 & 697  & 492 & 592 & 1073 & 1122  & 850 & 780 & 1776 & 1035\\
\hline
$0.8$ & 1175 & 994 & 699 & 864 & 1523 & 1591  & 1165 & 1073 & 2529 & 1444\\	
\hline	
		\end{tabular}
	}
	\caption{Big Mac data. Average asymmetric least squares loss $\delta_\uptau$ with leave-one-out cross validation is reported.}
	\label{bigmac}
\end{table}

\begin{figure}[t]
	\centering
	\includegraphics[width=5.5in,]{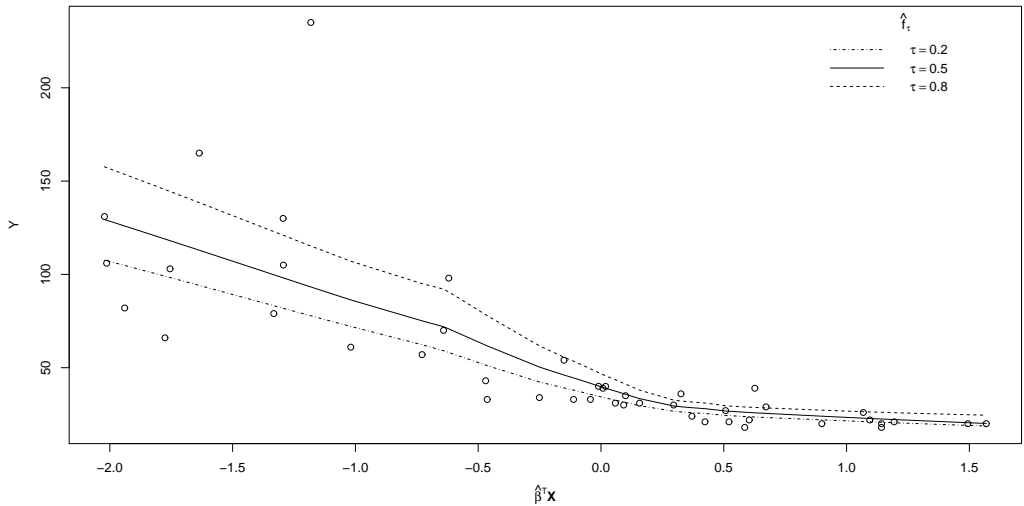}
	\caption{Scatter plot of the response $Y$ and the first sufficient direction $\hat{\bm \beta}^\top\bm{X}$ with fitted $\hat f_\uptau$ for $\uptau = 0.2$, $0.5$ and $0.8$ respectively.}
	\label{scatter}
\end{figure}

Next, we use  EA-SIR with $2$ slices to perform dimension reduction based on  the full data set. Denote the resulting central space estimator as $\hat{\bm{\beta}}$. For a fixed expectile level $\uptau$, we fit the kernel expectile regression between $Y$ and $\hat{\bm{\beta}}^\top\bm{X}$ based on the entire data, and denote the estimated $\uptau$-th conditional expectile function as $\hat f_\uptau$. For $\uptau=0.2$, $0.5$ and $0.8$, we plot $\hat f_\uptau$ versus $\hat{\bm{\beta}}^\top\bm{X}$ in Figure \ref{scatter}. We see a  monotone trend between $Y$ and $\hat{\bm{\beta}}^\top\bm{X}$, which explains the superior performance of EA-SIR over EA-DR in Table \ref{bigmac}. As we have seen in models IV and V
of the simulation studies, EA-SIR can improve over SIR in the presence of heteroscedasticity, which can be clearly seen from the estimated conditional expectile functions.

\section*{Acknowledgment}
The authors sincerely thank the associate editor and two anonymous reviewers for their comments that led to a much-improved paper. The authors thank Hyungwoo Kim for sharing the code of quantile-slicing mean estimation and  quantile-slicing variance estimation.

\section*{Appendix A: Proofs}
\noindent {\sc\bf Proof of Proposition \ref{proposition1}.} Let $\bm{B}$ be a basis of  $\mathcal{S}_{Y|\bm{X}}$.
Then we have 
$$E\big\{\phi_\uptau(Y - a) | \bm{X}=\bm{x} \big \}=E\big\{\phi_\uptau(Y - a) | \bm{B}^\top\bm{X}=\bm{B}^\top\bm{x} \big \}$$
because  $Y \indep \bm{X} | \bm{B}^\top\bm{ X} $.
By the definition in (\ref{als_loss}), $f_\uptau(\bm{X})$ evaluated at $\bm{x}$ becomes
 $$f_\uptau(\bm{x})=\argmin_a E\big\{\phi_\uptau(Y - a) | \bm{B}^\top\bm{X}=\bm{B}^\top\bm{x} \big \}.$$
 This implies that $f_\uptau(\bm{X})$ is a function of $\bm{B}^\top\bm{X}$ and $f_\uptau(\bm{X}) \indep \bm{X} | \bm{B}^ \top \bm{X}$ for any fixed $\uptau$.
 It follows that $\bm{\xi_X}\indep \bm{X} | \bm{B}^ \top \bm{X}$. By the definition of the central space, we have  $\mathcal{S}_{\bm{\xi_X} | \bm{X}}\subseteq  {\rm span}(\bm{B})  =\mathcal{S}_{Y|\bm{X}}$. \eop

Let $\bm{t} \in \mathbb{R}^k$ be a realization of $\bm{T}$. Denote $ \bm{\mu}_h(\bm{t})  = E\{\bm{Z} | \bm{\xi_X}^\top \bm{t} \in J_h(\bm{t})\}$.  The following Lemma is needed before we prove Proposition \ref{projectiveSIR}.

\begin{lemma}\label{lemma1}
Under Assumption \ref{linearity}, we have $\bm{\Sigma}^{-1/2}\bm{\mu}_h(\bm{t}) \in \mathcal{S}_{Y|\bm{X}}$.
\end{lemma}
\noindent {\sc\bf Proof of Lemma \ref{lemma1}.} Without loss of generality, assume $E(\bm{X})=\bm{0}$.  Let $\bm{B}$ be a basis of  $\mathcal{S}_{Y|\bm{X}}$. The LCM assumption implies that 
\begin{align}
E(\bm{X} | \bm{B}^\top{\bm{X}})=\bm{P}_{\bm{\Sigma}}(\bm{B}){\bm{X}},
    \label{21}
\end{align}
where $\bm{P}_{\bm{\Sigma}}(\bm{B})=\bm{\Sigma}\bm{B}(\bm{B}^\top\bm{\Sigma}\bm{B}) ^{-1}  \bm{B}^\top$. From the proof of Proposition \ref{proposition1}, we have $\bm{\xi_X}\indep \bm{X} | \bm{B}^ \top \bm{X}$, which implies that 
\begin{align}
\bm{\xi_X}^\top \bm{t}\indep \bm{X} | \bm{B}^ \top \bm{X}.
    \label{22}
\end{align}
Hence we have
 \begin{align}
 E(\bm{X}|\bm{\xi_X}^\top \bm{t})&=E\{E(\bm{X}|\bm{\xi_X}^\top \bm{t},\bm{B}^ \top \bm{X})|\bm{\xi_X}^\top \bm{t}\}=E\{E(\bm{X}|\bm{B}^ \top \bm{X})|\bm{\xi_X}^\top \bm{t}\},
 \label{23}
 \end{align}
where the first equality is due to the law of iterative expectations, and the second equality is guaranteed by  (\ref{22}). Plug (\ref{21}) into (\ref{23}) and we get
$$ E(\bm{X}|\bm{\xi_X}^\top \bm{t})= \bm{P}_{\bm{\Sigma}}(\bm{B}) E(\bm{X}|\bm{\xi_X}^\top \bm{t}).$$
It follows that
$$ \bm{\Sigma}^{-1}E(\bm{X}|\bm{\xi_X}^\top \bm{t})=\bm{B}(\bm{B}^\top\bm{\Sigma}\bm{B}) ^{-1}  \bm{B}^\top E(\bm{X}|\bm{\xi_X}^\top \bm{t})\subseteq  {\rm span}(\bm{B})  =\mathcal{S}_{Y|\bm{X}}.$$
Together with the fact that $\bm{Z} = \bm{\Sigma}^{-1/2}\bm{X}$ and the definition of $\bm{\mu}_h(\bm{t})$, we have
 $\bm{\Sigma}^{-1/2}\bm{\mu}_h(\bm{t}) \in \mathcal{S}_{Y|\bm{X}}$. \eop

\noindent {\sc\bf Proof of Proposition \ref{projectiveSIR}.} 
Let $\bm{t} \in \mathbb{R}^k$ be a realization of $\bm{T}$. Plug $\bm{T}=\bm{t}$ into (\ref{EASIR}) and we get
     $\bm{M}(\bm{t}) = \sum_{h=1}^H p_h(\bm{t}) \bm{\mu}_h(\bm{t}) \bm{\mu}_h^\top(\bm{t})$.
We know from Lemma \ref{lemma1} that ${\rm span}\{\bm{\Sigma}^{-1/2}\bm{M}(\bm{t})\}\subseteq  \mathcal{S}_{Y|\bm{X}}$.

 Let $\bm{\omega}\in \mathbb{R}^p$ belong to the orthogonal space of $\mathcal{S}_{Y|\bm{X}}$. Then it must also belong to the orthogonal space of ${\rm span}\{\bm{\Sigma}^{-1/2}\bm{M}(\bm{t})\}$. Thus we have $\bm{\omega}^\top\bm{\Sigma}^{-1/2}\bm{M}(\bm{t})=\bm{0}$ for all $\bm{t}$.
It follows that 
 \begin{align}
\bm{\omega}^\top\bm{\Sigma}^{-1/2}E\{\bm{M}(\bm{T})\}= E\{\bm{\omega}^\top\bm{\Sigma}^{-1/2}\bm{M}(\bm{T})\}=\bm{0}.
  \label{24}
 \end{align}
 Note that
 (\ref{24}) holds for any $\bm{\omega}$ orthogonal to $\mathcal{S}_{Y|\bm{X}}$. We have shown that
  \begin{align}
  \mathcal{S}_{Y|\bm{X}}^\perp\subseteq \left\{{\rm span}\big( \bm{\Sigma}^{-1/2} E\{ \bm{M}(\bm{T}) \} \big )\right\}^\perp,
    \label{25}
 \end{align}
 where $\perp$ denotes the orthogonal space. The conclusion follows from  (\ref{25}).\eop

Denote
$\bm{V}_h(\bm{t})=E\{\bm{Z}\bm{Z}^\top - \bm{I}_p | \bm{\xi_X}^\top \bm{t} \in J_h(\bm{t})\}$, where $\bm{t} \in \mathbb{R}^k$ be a realization of $\bm{T}$.
 The following Lemma is needed before we prove Proposition \ref{projectiveSAVE} and Proposition \ref{projectivedr}.

\begin{lemma}\label{lemma2}
Under Assumptions \ref{linearity} and \ref{ccv}, we have ${\rm span}\{\bm{\Sigma}^{-1/2}\bm{V}_h(\bm{t})\} \in \mathcal{S}_{Y|\bm{X}}$.
\end{lemma}

\noindent {\sc\bf Proof of Lemma \ref{lemma2}.} Without loss of generality, assume $E(\bm{X})=\bm{0}$.  Let $\bm{B}$ be a basis of  $\mathcal{S}_{Y|\bm{X}}$. The CCV assumption and the EV-VE formula lead to
 \begin{align}
 {\rm Var}(\bm{X}| \bm{B}^\top{\bm{X}})=E\{ {\rm Var}(\bm{X}| \bm{B}^\top{\bm{X}})\}=\bm{\Sigma}- {\rm Var}\{E(\bm{X}| \bm{B}^\top{\bm{X}})\}.
   \label{31}
 \end{align}
Recall from (\ref{21}) that $E(\bm{X} | \bm{B}^\top{\bm{X}})=\bm{P}_{\bm{\Sigma}}(\bm{B}){\bm{X}}$ under the LCM assumption. Together with 
(\ref{31}), we have
 \begin{align}
 E(\bm{X}\bm{X}^\top| \bm{B}^\top{\bm{X}})- \bm{P}_{\bm{\Sigma}}(\bm{B}){\bm{X}}{\bm{X}}^\top \bm{P}^\top_{\bm{\Sigma}}(\bm{B})=\bm{\Sigma}-\bm{P}_{\bm{\Sigma}}(\bm{B})\bm{\Sigma}\bm{P}^\top_{\bm{\Sigma}}(\bm{B}).
    \label{32}
 \end{align}
 After rearranging the terms of (\ref{32}), we have
  \begin{align}
 E\{(\bm{X}\bm{X}^\top-\bm{\Sigma})| \bm{B}^\top{\bm{X}}\}=\bm{P}_{\bm{\Sigma}}(\bm{B})(\bm{X}\bm{X}^\top-\bm{\Sigma})\bm{P}^\top_{\bm{\Sigma}}(\bm{B}).
    \label{33}
 \end{align}
 Note that
  \begin{align}
  \begin{split}
 E
 \{(\bm{X}\bm{X}^\top-\bm{\Sigma})|\bm{\xi_X}^\top \bm{t}\}&=E[E\{(\bm{X}\bm{X}^\top-\bm{\Sigma})|\bm{\xi_X}^\top \bm{t},\bm{B}^ \top \bm{X}\}|\bm{\xi_X}^\top \bm{t}]\\
 &=E[E\{(\bm{X}\bm{X}^\top-\bm{\Sigma})|\bm{B}^ \top \bm{X}\}|\bm{\xi_X}^\top \bm{t}],
 \end{split}
 \label{34}
  \end{align}
  where the first equality is due to the law of iterative expectations, and the second equality is guaranteed by  (\ref{22}).  Plug (\ref{33}) into (\ref{34}) and we get
  \begin{align}
E
 \{(\bm{X}\bm{X}^\top-\bm{\Sigma})|\bm{\xi_X}^\top \bm{t}\}=
\bm{P}_{\bm{\Sigma}}(\bm{B}) E
 \{(\bm{X}\bm{X}^\top-\bm{\Sigma})|\bm{\xi_X}^\top \bm{t}\}\bm{P}^\top_{\bm{\Sigma}}(\bm{B}).
    \label{35}
 \end{align}
Recall that
  $\bm{Z} = \bm{\Sigma}^{-1/2}\bm{X}$
 and
  $\bm{P}_{\bm{\Sigma}}(\bm{B})=\bm{\Sigma}\bm{B}(\bm{B}^\top\bm{\Sigma}\bm{B}) ^{-1}  \bm{B}^\top$.
 It follows from (\ref{35}) that
  \begin{align*}
  &\bm{\Sigma}^{-1/2}E\{(\bm{Z}\bm{Z}^\top - \bm{I}_p) | \bm{\xi_X}^\top \bm{t}\}\bm{\Sigma}^{-1/2} =
  \bm{\Sigma}^{-1}E
 \{(\bm{X}\bm{X}^\top-\bm{\Sigma})|\bm{\xi_X}^\top \bm{t}\} \bm{\Sigma}^{-1}\\&\hspace{.2in}=\bm{B}(\bm{B}^\top\bm{\Sigma}\bm{B}) ^{-1}  \bm{B}^\top E
 \{(\bm{X}\bm{X}^\top-\bm{\Sigma})|\bm{\xi_X}^\top \bm{t}\} \bm{B}(\bm{B}^\top\bm{\Sigma}\bm{B}) ^{-1}  \bm{B}^\top \\
 &\hspace{.2in} \subseteq  {\rm span}(\bm{B})  =\mathcal{S}_{Y|\bm{X}}.
  \end{align*}
  Together with the definition of $\bm{V}_h(\bm{t})$, we have shown ${\rm span}\{\bm{\Sigma}^{-1/2}\bm{V}_h(\bm{t})\} \in \mathcal{S}_{Y|\bm{X}}$.\eop

\noindent {\sc\bf Proof of Proposition \ref{projectiveSAVE}.} 
Let $\bm{G}(\bm{t})$ be a realization of $\bm{G}(\bm{T})$. From Lemma \ref{lemma1}, Lemma \ref{lemma2} and the definition of $\bm{G}(\bm{T})$, we have ${\rm span}\{\bm{\Sigma}^{-1/2}\bm{G}(\bm{t})\}\subseteq  \mathcal{S}_{Y|\bm{X}}$. The rest of the proof is exactly parallel to the proof of Proposition \ref{projectiveSIR},  and is thus omitted.
\eop

\noindent {\sc\bf Proof of Proposition \ref{projectivedr}.} Let $\bm{F}(\bm{t})$ be a realization of $\bm{F}(\bm{T})$. From Lemma \ref{lemma1}, Lemma \ref{lemma2} and the definition of $\bm{F}(\bm{T})$, we have ${\rm span}\{\bm{\Sigma}^{-1/2}\bm{F}(\bm{t})\}\subseteq  \mathcal{S}_{Y|\bm{X}}$. The rest of the proof is exactly parallel to the proof of Proposition \ref{projectiveSIR},  and is thus omitted.
\eop

\noindent {\sc\bf Proof of Proposition \ref{msir}.} Let $\bm{B}$ be a basis of  $\mathcal{S}_{Y|\bm{X}}$.
From the proof of Proposition \ref{proposition1}, we have $f_{\uptau_\ell}(\bm{X})\indep \bm{X} | \bm{B}^ \top \bm{X}$. Similar to the proof of Lemma \ref{lemma1}, we can show that
$$ E\{\bm{X}|f_{\uptau_\ell}(\bm{X})\}=\bm{\Sigma}\bm{B}(\bm{B}^\top\bm{\Sigma}\bm{B}) ^{-1}  \bm{B}^\top E\{\bm{X}|f_{\uptau_\ell}(\bm{X})\}.$$
Thus we have $\bm{\Sigma}^{-1} E\{\bm{X}|f_{\uptau_\ell}(\bm{X})\}\in  {\rm span}(\bm{B})  =\mathcal{S}_{Y|\bm{X}}$. From the definition of $\bm{M}_\ell$, it follows that ${\rm span}(\bm{\Sigma}^{-1/2} \bm{M}_\ell)\subseteq \mathcal{S}_{Y|\bm{X}}$ for $\ell=1,\ldots,k$.
Hence we have
${\rm span}(\bm{\Sigma}^{-1/2} \widetilde{\bm{M}})={\rm span}(\bm{\Sigma}^{-1/2} \bm{M}_1,\ldots,\bm{\Sigma}^{-1/2} \bm{M}_k)\subseteq \mathcal{S}_{Y|\bm{X}}$.
\eop

\section*{Appendix B: Additional simulation results}

We present additional simulation results in this Appendix to inspect the effects of $H$, $N$, $k$, $r$ and $\lambda$ for expectile-assisted estimators. From Table \ref{effect_H}, we see that EA-SIR and EA-DR are not very sensitive to the choice of number of slices in both model I and model II. EA-SAVE has stable performance in model II, but becomes worse in model I when $H$ increases from $4$ to $10$.

\begin{table}[H]
\captionsetup{font=small}
\centering
\begin{minipage}{0.95\textwidth}
	\begin{tabular}{ccccc}
\hline
		Model & Method &  $H=4$ &  $H=5$  &  $H=10$   \\ \hline 
				\multirow{3}{*}{I}  & EA-SIR & 1.364 (0.060)  &  1.343   (0.058) &  1.217 (0.060)  \\

		& EA-SAVE  &  0.489  (0.044)   & 0.554 (0.050)	  &   1.228 (0.062) \\
		
		& EA-DR &  0.365 (0.036) & 0.345 (0.029)    &   0.380 (0.035)  \\\hline
						\multirow{3}{*}{II}  & EA-SIR & 1.525 (0.046)   &   1.567  (0.046) &  1.497 (0.047)   \\
			& EA-SAVE  &  1.515  (0.051)	  &  1.543  (0.048)  &  1.616   (0.041)  \\
				& EA-DR &  1.496 (0.049) &  1.497  (0.050) &  1.517 (0.049)  \\\hline		
	\end{tabular}	\caption{Effect of $H$ (number of slices). The average of $\Delta$ and its standard error (in parentheses) are reported based on 100 repetitions for $(n,p)=(100,6)$. }
	\label{effect_H}
	\end{minipage}
	\end{table}

\begin{table}[H]
\captionsetup{font=small}
\centering
\begin{minipage}{1\textwidth}
	\scalebox{.9}{ 
	\begin{tabular}{cccccc}
\hline
		Model & Method &  $N=100$ &  $N=200$  &  $N=500$  &  $N=1000$   \\ \hline 
				\multirow{3}{*}{I}  & EA-SIR &  1.464 (0.052)	& 1.444 (0.052)  & 1.431 (0.052) 	& 1.343 (0.058)  \\

		& EA-SAVE  & 0.584 (0.054)	& 0.546 (0.052)	& 0.558  (0.053)	& 0.554	 (0.050)\\
		
		& EA-DR & 0.359 (0.033) & 0.359 (0.033) & 0.356 (0.033) &  0.345 (0.029) \\ \hline

		\multirow{3}{*}{II}  & EA-SIR & 1.494 (0.047)  & 1.519 (0.047)  & 1.505 (0.048)  & 1.567 (0.046)   \\
		
		& EA-SAVE  & 1.514 (0.049) & 1.495 (0.048) & 1.502 (0.047) & 1.543 (0.048)  \\
		
		& EA-DR & 1.497 (0.051) & 1.472 (0.052) & 1.461 (0.053) & 1.497 (0.050)  \\ \hline
		
	\end{tabular}}
	\caption{Effect of $N$ (number of projections). The average of $\Delta$ and its standard error (in parentheses) are reported based on 100 repetitions for $(n,p)=(100,6)$. }
	\label{effect_N}
	\end{minipage}
\end{table}

	From    Table \ref{effect_N}, we see that all three expectile-assisted methods are not overly sensitive to $N$, which denotes the number of projections. 

Table \ref{effect_k} summarizes the results for different $k$. For $k=4$, we set the expectile levels to be $0.2$, $0.4$, $0.6$ and $0.8$; for $k=9$, we set the expectile levels to be $0.1,0.2,\ldots,0.9$; for $k=19$, the expectile levels are $0.05,0.1,0.15,\ldots,0.95$. We see from Table \ref{effect_k} that the expectile-assisted methods are not very sensitive to the choice of different expectile levels. 

	Table \ref{effect_r} summarizes the results for different $r$ in the Gaussian radial basis kernel (\ref{rbf}). We see that the results are stable when $r$ varies around  $1/\gamma^2$, which is the suggested value in (\ref{r value}).
	
	 We compare the choice of fixed $\lambda$ versus the data-driven $\lambda$ in Table \ref{effect_lambda}. Please refer to Section 5.1 for the details of the data-driven approach to choose $\lambda$, which is in the last column of Table \ref{effect_lambda}. The best $\lambda$ that corresponds to the optimal result (boldfaced for easy reference)  changes across different models and different methods, and the data-driven $\lambda$ always achieves a decent result that is very close to the optimal result.

	\begin{table}[H]
\captionsetup{font=small}
\centering
\begin{minipage}{0.95\textwidth}
	\begin{tabular}{ccccc}
\hline
		Model & Method &  $k=4$ &  $k=9$  &  $k=19$   \\ \hline 
				\multirow{3}{*}{I}  & EA-SIR & 1.455 (0.052)  &  1.343   (0.058) &  1.429 (0.054)  \\

		& EA-SAVE  &  0.567  (0.052)   & 0.554 (0.050)	  &   0.549 (0.051) \\
		
		& EA-DR &  0.358 (0.034) & 0.345 (0.029)    &   0.368 (0.035)  \\\hline
						\multirow{3}{*}{II}  & EA-SIR & 1.504 (0.049)   &   1.567  (0.046) &  1.512 (0.047)   \\
			& EA-SAVE  &  1.489  (0.050)	  &  1.543  (0.048)  &  1.487   (0.049)  \\
				& EA-DR &  1.472 (0.053) &  1.497  (0.050) &  1.452 (0.053)  \\\hline		
	\end{tabular}	
	\caption{Effect of  $k$ (number of expectile levels). The average of $\Delta$ and its standard error (in parentheses) are reported based on 100 repetitions for $(n,p)=(100,6)$. }
	\label{effect_k}
	\end{minipage}
	\end{table}

\begin{table}[H]
\captionsetup{font=small}
\centering
\begin{minipage}{0.95\textwidth}
	\begin{tabular}{ccccccc}
\hline
		\multirow{2}{*}{Model} & \multirow{2}{*}{Method} & \multicolumn{5}{c}{$r$}   \\ \cline{3-7}
		&  & $1/(4\gamma^2)$  &  $1/(2\gamma^2)$ & $1/\gamma^2$ & $2/\gamma^2$ & $4/\gamma^2$  \\ \hline 
				\multirow{6}{*}{I}  & \multirow{2}{*}{EA-SIR} &  1.294	& 1.308 & 1.343	& 1.344 & 1.431\\
		&   &  (0.058) & (0.058) & (0.058) & (0.056) & (0.053)\\ 
		\cline{3-7}
		
		& \multirow{2}{*}{EA-SAVE}  & 0.571	& 0.544	& 0.554	& 0.576	& 0.541\\
		&  & (0.045) & (0.047)	& (0.050)	& (0.051)	& (0.051) \\
		\cline{3-7}
		
		& \multirow{2}{*}{EA-DR} & 0.463 & 0.412 & 0.345 & 0.368 & 0.364 \\
		&  & (0.035) & (0.037) & (0.029) & (0.033) & (0.034) \\ \hline

		\multirow{6}{*}{II}  & \multirow{2}{*}{EA-SIR} & 1.601 & 1.567 & 1.567 & 1.507 & 1.508 \\
		&   &  (0.042) & (0.042) & (0.046)	& (0.047)	& (0.047) \\ 
		\cline{3-7}
		
		& \multirow{2}{*}{EA-SAVE}  & 1.562 & 1.535 & 1.543 & 1.556 & 1.494 \\
		&  &  (0.048) & (0.051) & (0.048) & (0.047) & (0.046) \\
		\cline{3-7}
		
		& \multirow{2}{*}{EA-DR} & 1.602 & 1.577 & 1.497 & 1.504 & 1.476 \\
		&  & (0.044) & (0.046) & (0.050) & (0.049) & (0.050) \\ \hline
		
	\end{tabular}
	\caption{Effect of $r$ in  (\ref{rbf}). The average of $\Delta$ and its standard error (in parentheses) are reported based on 100 repetitions for $(n,p)=(100,6)$. }
	\label{effect_r}
	\end{minipage}
\end{table}

\section*{Appendix C: Order determination}

We take a sequential test approach for order determination. 
Consider
\begin{align}
H_0^{(m)}:d=m\mbox{ v.s. }H_a^{(m)}:d>m,\mbox{ for }m=0,1,\ldots,p-1.
  \label{order1}
 \end{align}
Then we estimate the structural dimension $d$ by the smallest $m$ at which 
$H_0^{(m)}$ in (\ref{order1}) is accepted. The asymptotic sequential tests for SIR, SAVE and directional regression are well-known in the literature. See, for example, Chapter 9 of Li (2018).

\begin{table}[t]
\captionsetup{font=small}
\centering
\begin{minipage}{0.99\textwidth}
	\begin{tabular}{cccccccc}
		\hline
		\multirow{2}{*}{Model} & \multirow{2}{*}{Method} & \multicolumn{5}{c}{$\lambda$}   & \multirow{2}{*}{$\textrm{DD}$} \\ \cline{3-7}
		&  & 0.001 & 0.01 & 0.1 & 1 & 10 &  \\ \hline 
		
		\multirow{6}{*}{I}  & \multirow{2}{*}{EA-SIR} & 1.675 &		1.67 & 1.634 & 1.524 &  {\bf 1.339} &   1.343\\
		&   &  (0.042) & (0.044) & (0.043) & (0.050) & (0.062) &  (0.058)\\ 
		\cline{3-8}
		& \multirow{2}{*}{EA-SAVE}  & 0.643	& 0.619	& {\bf 0.576} & 0.927	& 1.261 &  0.554\\
		&  &  (0.056) & (0.054)	& (0.047) & (0.058)	& (0.058) &  (0.050)\\
		\cline{3-8}
		& \multirow{2}{*}{EA-DR} & {\bf 0.357} & 0.376 & 0.486 & 0.904	& 1.114 &  0.345\\
		&  &  (0.033) & (0.034)	& (0.041) & (0.053)	& (0.06) &  (0.029) \\ \hline
		
		\multirow{6}{*}{II}  & \multirow{2}{*}{EA-SIR} &  {\bf 1.521} & 1.589 & 1.608	 & 1.576 & 1.662 &  1.567\\
		&   & (0.047) & (0.041) & (0.040) & (0.044) & (0.044) &  (0.046)\\ 
		\cline{3-8}
		& \multirow{2}{*}{EA-SAVE}  & 1.573	&  {\bf 1.558}	& 1.574	& 1.600 & 1.599 &  1.543\\
		&  & (0.046) & (0.047)	 & (0.047)	& (0.045)	& (0.047) &  (0.048) \\
		\cline{3-8}
		& \multirow{2}{*}{EA-DR} &  {\bf 1.492} & 1.517 & 1.574 & 1.645 & 1.647 &  1.497 \\
		&  & (0.050)	& (0.047)	& (0.045)	& (0.041)	& (0.042) &  (0.050)
		\\ \hline
	\end{tabular}
	\caption{Fixed $\lambda$ in (\ref{obj1}) versus data-driven  $\lambda$ (DD). The average of $\Delta$ and its standard error (in parentheses) are reported based on 100 repetitions for $(n,p)=(100,6)$. }
	\label{effect_lambda}
		\end{minipage}
\end{table}

Next, we describe a permutation test approach for order determination based on  EA-SIR. Given an i.i.d. sample $\{(\bm{X}_i,Y_i):i=1,\ldots, n\}$, let $\hat{\bm{M}}(\bm{T})$ denote  the sample estimator of $E\{ \bm{M}(\bm{T}) \}$. Suppose $\hat \eta_1\ge \hat\eta_2\ge\ldots\ge \hat \eta_p$ are the eigenvalues of $\hat{\bm{M}}(\bm{T})$, and the corresponding eigenvectors are $\hat{\bm{v}}_1,..,\hat{\bm{v}}_p$ . Consider test statistic
\begin{align}
\Lambda_m=n\sum_{j=m+1}^p \hat \eta_j.
  \label{order2}
 \end{align}
Following Section 3.3 of Cook and Yin (2001), we have the following algorithm
\begin{enumerate}
\item Calculate standardized predictors $\bm{\hat{Z}}_i = \hat{\bm{\Sigma}}^{-1/2}(\bm{X}_i-\hat{\bm{\mu}})$ for $i=1,\ldots,n$, where $\hat{\bm{\mu}}=n^{-1}\sum_{i=1}^n \bm{X}_i$ and $\hat{\bm{\Sigma}}=n^{-1}\sum_{i=1}^n ( \bm{X}_i-\hat{\bm{\mu}})( \bm{X}_i-\hat{\bm{\mu}})^\top$.  
     \item Denote $\hat{\bm{U}}_1=(\hat{\bm{v}}_1,..,\hat{\bm{v}}_m)$
     and $\hat{\bm{U}}_2=(\hat{\bm{v}}_{m+1},..,\hat{\bm{v}}_p)$. Construct sample principal predictors $\hat{\bm{W}}_{1i}=\hat{\bm{U}}_1^\top \bm{\hat{Z}}_i$ and $\hat{\bm{W}}_{2i}=\hat{\bm{U}}_2^\top \bm{\hat{Z}}_i$, $i=1,\ldots, n$.
     \item For $b = 1,\ldots, B$, randomly permute the indices $i$ of  $\hat{\bm{W}}_{2i}$ to obtain the permuted set  $\hat{\bm{W}}_{2i}^{[b]}$. Construct the test statistic $\Lambda_m^{[b]}$ in (\ref{order2}) based on $\{(\hat{\bm{W}}_{1i},\hat{\bm{W}}_{2i}^{[b]},Y_i): i=1,\ldots, n\}$.
     \item Calculate the p-value as $p^{(m)}=B^{-1}\sum_{b=1}^B I(\Lambda_m^{[b]}>\Lambda_m)$. For a prespecified significance level $\alpha$, reject  $H_0^{(m)}$ in (\ref{order1}) if $p^{(m)}<\alpha$.
\end{enumerate}

\noindent A similar permutation test approach can be implemented for EA-SAVE, EA-DR, and their corresponding marginal expectile-assisted methods.
When we combine the projection step in the algorithm from Section 3.3 with the permutation step 3 above, the computation becomes very time-consuming. The marginal expectile-assisted methods in Section 5.2 are computationally more efficient as no projection is needed.

 Denote $\hat d$ as the estimated structural dimension. We report the frequency of $\hat d$ in Table \ref{order} based on $100$ repetitions. We fix $p=6$ and set $n=100$ or $300$. Directional regression is used for the asymptotic  test, and marginal EA-DR with  $B=200$ permutations is used for the permutation test.  The significance level is set as $\alpha=0.1$. The true structural dimension is $d=2$ for model I and $d=1$ for model V. As $n$ increases, both the asymptotic test and the permutation test lead to higher frequency of correct estimation. While both tests work well for model I, the asymptotic test is better than the permutation test for model V.

\begin{table}[t]
\captionsetup{font=small}
\centering
\begin{minipage}{0.99\textwidth}
	\begin{tabular}{c|c|cccc|cccc}
	\hline
	\multirow{2}{*}{Model} & \multirow{2}{*}{$n$} & \multicolumn{4}{c|}{Asymptotic} & \multicolumn{4}{c}{Permutation}\\
	& & $\hat d=0$ & $\hat d=1$ & $\hat d=2$ & $\hat d>2$   & $\hat d=0$ & $\hat d=1$ & $\hat d=2$ & $\hat d>2$  \\\hline
	 \multirow{2}{*}{I} & 100 & 0& 28& 59 & 13 & 0 & 17 & 71 & 12\\
	 & 300 & 0 & 0 & 96 & 4 &  0 & 0 & 89 & 11\\
	 \hline
	 	 \multirow{2}{*}{V} &  100 & 8 & 82& 9& 1 & 55 & 38 & 6 & 1\\
	 & 300 & 0 & 89 & 10 & 1 &  49 & 43 & 5 & 3\\
	 \hline
	\end{tabular}
	\caption{Sequential test for order determination with $\alpha=0.1$. The frequency of $\hat d$ is reported based on 100 repetitions for $p=6$. }
	\label{order}
		\end{minipage}
\end{table}


\begin{thebibliography}{99}
	
\bibitem{} 
Barreda, L., Gannoun, A. and Saracco, J. (2007)
Some extensions of multivariate sliced inverse regression. 
{\it Journal of Statistical Computation and Simulation}, {\bf 77}, 1--17.	
	
	\bibitem{}
	Box, G. E. and Cox, D. R. (1964) An analysis of transformations. {\it Journal of the Royal Statistical Society: Series B}, {\bf 26}, 211--252.	
	
		\bibitem{}
	Cai, Z., Li, R. and Zhu, L. (2020) Online sufficient dimension reduction through sliced inverse regression. {\it Journal of Machine Learning Research}, {\bf 21(10)}, 1--25.
		
		\bibitem{}
 Chavent, M., Girard, S., Kuentz-Simonet, V.,  Liquet, B., Nguyen, T. and  Saracco, J. (2014) A sliced inverse regression approach for data stream. {\it Computational Statistics}, {\bf 29}, 1129--1152. 

	
\bibitem{}
Chen, J. (2018) On exactitude in financial regulation: value-at-risk, expected shortfall, and expectiles. {\it Risks}, {\bf 6}, 61.


\bibitem{}
Chen, X., Zou, C. and Cook, R. D. (2010) Coordinate-independent sparse sufficient dimension reduction and variable selection. {\it The Annals of Statistics}, {\bf 38}, 3696--3723.

\bibitem{Cook1998}
Cook, R. D. (1998a) {\it Regression graphics: ideas for
	studying regressions through graphics}. New York: Wiley.

\bibitem{}
Cook, R. D. (1998b) Principal Hessian directions revisited. {\it Journal of the American Statistical Association}, {\bf 93}, 84--94.

\bibitem{}
Cook, R. D. (2007) Fisher lecture: dimension reduction in regression. {\it Statistical Science}, {\bf 22}, 1--26.

\bibitem{}
Cook, R. D. and Setodji, M. (2003) A model-free test for reduced rank in multivariate regression. \textit{Journal of the American Statistical Association}, {\bf 98}, 340--351.

\bibitem{}
Cook, R. D. and  Weisberg, S. (1991)  Comment on ``Sliced inverse regression for dimension reduction''. \textit{ Journal of American Statistical Association}, {\bf 86}, 28--33.

\bibitem{}
Cook, R. D. and Yin, X. (2001) Dimension reduction and visualization  in discriminant analysis  (with discussion). {\it Australian and New Zealand Journal of Statistics}, {\bf 43}, 147--199.
 
\bibitem{}
Coudret, R., Girard, S. and Saracco, J. (2014) A new sliced inverse regression method for multivariate response. Computational Statistics and Data Analysis, {\bf 77}, 285--299. 

\bibitem{}
Daouia, A., Girard, S. and Stupfler, G. (2018) Estimation of tail risk based on extreme expectiles. {\it Journal of the Royal Statistical Society: Series B}, {\bf 80}, 263--292.

\bibitem{}
Dong, Y. (2020) A brief review of linear sufficient dimension reduction through optimization. To appear in {\it Journal of Statistical Planning and Inference}.


\bibitem{}
Guo, Z., Li, L.,  Lu, W. and  Li, B. (2015) Groupwise dimension reduction via envelope method. {\it Journal of the American Statistical Association},  {\bf 110}, 1515--1527.


\bibitem{}
Kim, H., Wu, Y. and Shin, S. J. (2019) Quantile-slicing estimation for dimension reduction in regression. {\it Journal of Statistical Planning and Inference}, {\bf 198}, 1--12.

\bibitem{}
Kim, K., Li, B., Yu, Z. and Li, L. (2020) On post dimension reduction statistical inference. To appear in  {\it The Annals of Statistics}.

\bibitem{}
Kim, M. and Lee, S. (2016)
Nonlinear expectile regression with application to value-at-risk and expected shortfall estimation. {\it Computational Statistics and Data Analysis}, {\bf 94}, 1--19.


\bibitem{}
 Lee, K. Y.,  Li, B. and Chiaromonte, F. (2013) A general theory for nonlinear sufficient dimension reduction: formulation and estimation. {\it The Annals of Statistics}, {\bf 41}, 221--249.




\bibitem{}
Li, B. (2018)
{\it Sufficient dimension reduction: methods and applications with R}.
CRC Press.

\bibitem{}
Li, B., Artemiou, A. and Li, L. (2011) Principal support vector machines for linear and nonlinear sufficient dimension reduction. {\it  The Annals of Statistics}, {\bf 39}, 3182--3210.

\bibitem{}
Li, B. and Wang, S. (2007) On directional regression for dimension reduction. \textit{Journal of American Statistical Association},  {\bf 479}, 997--1008.

\bibitem{}
Li, B., Wen, S. and Zhu, L. X. (2008) On a projective resampling method
for dimension reduction with multivariate responses. \textit{Journal of American Statistical Association}, {\bf 103}, 1177--1186.

\bibitem{} Li, K. C. (1991) Sliced inverse regression for dimension
reduction (with discussion). {\it Journal of the American
	Statistical Association}, {\bf 86}, 316--342.

\bibitem{}
Li, K. C. (1992) On principal Hessian directions for data visualization and dimension reduction: another application of Stein's lemma. {\it Journal of the American Statistical Association}, {\bf 87}, 1025--1039.


\bibitem{}
Li, L. (2007) Sparse sufficient dimension reduction. {\it Biometrika}, {\bf 94}, 603--613.


\bibitem{}
Li, L., Li, B. and Zhu, L. X. (2010) Groupwise dimension reduction.  \textit{Journal of American Statistical Association}, {\bf 105}, 1188--1201.


\bibitem{}
Li, Y. and Zhu, L. X. (2007) Asymptotics for sliced inverse variance estimation. {\it The Annals of Statistics}, {\bf 35}, 41--69.

\bibitem{}
Luo, W., Li, B. and Yin, X. (2014) On efficient dimension reduction with respect to a statistical functional of interest. {\it The Annals of Statistics}, {\bf 42}, 382--412.

\bibitem{}
Ma, Y. and Zhu, L. (2012) A semiparametric approach to dimension reduction. {\it Journal of the American Statistical Association}, {\bf 107}, 168--179.


\bibitem{}
Ma, Y. and Zhu, L. (2013) 
A review on dimension reduction.   {\it International Statistics Review}, {\bf 81}, 134--150. 


\bibitem{}
Newey, W. K. and  Powell, J. L. (1987) Asymmetric least squares estimation and testing. {\it Econometrica}, {\bf 55}, 819--847.




\bibitem{}
Saracco, J. (2005) Asymptotics for pooled marginal slicing estimator based on $\rm{SIR}_\alpha$ approach. {\it Journal of Multivariate Analysis}, {\bf  96}, 117--135.

\bibitem{}
Sz{\'e}kely, G. J., Rizzo, M. L. and  Bakirov, N. K.  (2007) Measuring and testing dependence by correlation of distances. {\it The Annals of Statistics}, {\bf 35}, 2769--2794.


\bibitem{}
Tan, K., Shi, L. and Yu, Z. (2020) Sparse SIR: optimal rates and adaptive estimation.  {\it The Annals of Statistics}, {\bf 48}, 64--85.


\bibitem{}
Wang, C.,  Shin, S. J.  and Wu, Y. (2018) Principal quantile regression for sufficient dimension reduction with heteroscedasticity. {\it Electronic Journal of Statistics}, {\bf 12}, 2114--2140.

\bibitem{}
Wang, Q. and Yin, X. (2008) A nonlinear multi-dimensional variable selection method for high dimensional data: sparse MAVE. {\it Computational Statistics and Data Analysis}, {\bf 52}, 4512--4520.


\bibitem{} Xia, Y., Tong, H., Li, W. K. and Zhu, L. X. (2002) An adaptive estimation of optimal regression subspace. {\it Journal of the Royal Statistical Society: Series B}, {\bf 64}, 363--410.

\bibitem{}
Yang, Y., Zhang, T. and Zou, H. (2018) Flexible expectile regression in reproducing kernel Hilbert spaces. {\it Technometrics}, {\bf 60}, 26--35.

\bibitem{}
Yao, Q. and Tong, H. (1996) Asymmetric least squares regression estimation: a nonparametric approach. {\it Journal of Nonparametric Statistics}, {\bf 6},
273--292.

\bibitem{}
Yin, X. and Bura, E. (2006)
Moment-based dimension reduction for multivariate response regression. {\it Journal of Statistical Planning and Inference}, {\bf 136}, 3675--3688.

\bibitem{}
Yin, X. and Cook, R. D. (2003) Estimating central subspaces via inverse third moments. {\it Biometrika}, {\bf 90}, 113--125.

\bibitem{}
Yin, X. and Li, B. (2011) Sufficient dimension reduction based on an ensemble of minimum average variance estimators. {\it The Annals of Statistics}, {\bf 39}, 3392--3416.

\bibitem{}
Yin, X., Li, B. and Cook, R. D. (2008) Successive direction extraction for estimating the central subspace in a multiple-index regression. {\it Journal of Multivariate Analysis}, {\bf 99}, 1733--1757.

\bibitem{}
Yu, Z.,  Zhu, L.,  Peng, H. and  Zhu, L. X. (2013) Dimension reduction and predictor selection in semiparametric models. {\it Biometrika}, {\bf 100}, 641--654.


\bibitem{}
Zhu, L. X. and Ng, K. W. (1995) Asymptotics of sliced inverse regression. {\it Statistica Sinica}, {\bf 5}, 727--736.






























\end{thebibliography}
\end{document}